\documentclass[preprint, amsmath, amssymb, showkeys, titlepage, superscriptaddress]{revtex4-2}

\setcitestyle{authoryear,round}
\usepackage{graphicx}
\usepackage[bookmarks=true, colorlinks=true, citecolor=black, linkcolor=black, urlcolor=black]{hyperref}
\usepackage{booktabs}
\usepackage{rotating}

\def\sym#1{\ifmmode^{#1}\else\(^{#1}\)\fi}

\begin{document}

\title{Relying on recent and temporally dispersed science predicts breakthrough inventions}

\author{Qing Ke}
\affiliation{Department of Data Science, College of Computing, City University of Hong Kong, Hong Kong, China}

\author{Ziyou Teng}
\affiliation{Department of Data Science, College of Computing, City University of Hong Kong, Hong Kong, China}
\affiliation{Shanghai International College of Intellectual Property, Tongji University, Shanghai, China}

\author{Chao Min}
\email{mc@nju.edu.cn}
\affiliation{School of Information Management, Nanjing University, Nanjing, China}

\date{\today}

\begin{abstract}
The development of inventions is theorized as a process of searching and recombining existing knowledge components. Previous studies under this theory have examined myriad characteristics of recombined knowledge and their performance implications. One such feature that has received much attention is technological knowledge age. Yet, little is known about how the age of scientific knowledge influences the impact of inventions, despite the widely known catalyzing role of science in the creation of new technologies. Here we use a large corpus of patents and derive features characterizing how patents temporally search in the scientific space. We find that patents that cite scientific papers have more citations and substantially more likely to become breakthroughs. Conditional on searching in the scientific space, referencing more recent papers increases the impact of patents and the likelihood of being breakthroughs. However, this positive effect can be offset if patents cite papers whose ages exhibit a low variance. These effects are consistent across technological fields. 
\end{abstract}


\maketitle

\section{Introduction}

In today's knowledge-based economy, innovation has been playing an important role in productivity growth and competitive advantage of firms~\citep{Ashish2018decline, Fleming2004science, nelson1982role}. In the process of the production of innovation, creators---scientists and inventors---face an enormously large space of knowledge from which they can recombine existing knowledge components as input~\citep{Lee2001recombinant, savino2017search}. The questions of how they search in the knowledge landscape and how the search process is linked to the value of resultant innovation have received a considerate amount of attention, with myriad types of searches being theorized and examined, including local search~\citep{rosenkopf2001beyond}, broad search~\citep{katila2002something, leiponen2010innovation}, repeated search~\citep{katila2002something}, originality search~\citep{jung2016quest}, among many others. In addition, characteristics of recombined components have also been explored, including their locations in technological and geographical spaces~\citep{phene2006breakthrough}, organizational context~\citep{yang2010learning}, reuse trajectories~\citep{kok2019dusting, wang2014knowledge}, etc. 

This study concerns about temporal search in the context of inventions: Inventors may recombine prior knowledge that are produced at different points in time, and the extent to which recent versus mature knowledge is recombined may have profound implications for the value of new inventions~\citep{katila2002new}. Consequently, the costs and benefits of both recent and mature knowledge have been extensively discussed. On one hand, recent knowledge has been emphasized as valuable, as it may open up avenues for new ideas and practices. The space around recent knowledge provides ample recombinant opportunities, leaving more rooms for explorations of new ideas for valuable inventions~\citep{heeley2008recency}. Evidence supporting these benefits indicates that firms experimenting with emerging technologies can overcome the ``maturity trap'' and create breakthrough inventions~\citep{ahuja2001entrepreneurship} and combining recent knowledge with knowledge with large time spans is associated with the impact of inventions~\citep{nerkar2003old}. 

Against this so-called ``recency bias'', mature knowledge has also been advocated to be useful to inventions for several reasons. First, it may reduce uncertainty and potential technical errors in applications, making it more reliable than newly created knowledge~\citep{Lee2001recombinant, petruzzelli2014search}. Second, mature knowledge may reduce its utilization costs due to its compatibility with existing knowledge, making it easy to be integrated into new applications~\citep{petruzzelli2014search, turner2013strategic}. Third, a certain level of maturity is often needed to make innovation acceptable, as new ideas without connections to current knowledge can encounter resistance to be recognized~\citep{Petruzzelli2018maturity}. Furthermore, the importance of some types of knowledge can only be appreciated after sufficient amount of time when discoveries therein are brought to light due to factors like the advancement of enabling technologies~\citep{cattani2006technological, nerkar2003old}. However, the cost of overly mature knowledge is that it may obsolete and lose its relevance over time. Moreover, old knowledge may hinder individual and organizational learning~\citep{ahuja2001entrepreneurship, katila2002new}, and inventors who continuously use such knowledge may fall into ``competency traps''~\citep{levitt1988organizational}, meaning that they are not able to learn superior new knowledge and practices and innovation value generated by past knowledge markedly decreases. 

Despite the relevance of temporality to the search process of inventions, the knowledge space focused in previous discussions has been limited exclusively to the technological space. Inventions, however, rely on not only technological knowledge but also scientific one, and the role of science in the development of new technologies has long been substantiated. Science may alter inventors' search process towards more useful recombinations~\citep{Fleming2004science}; inventions that are closer to science through citation relations are more impactful and valuable~\citep{Ahmadpoor2017dual}; the quality of scientific papers referenced by patents has a strong positive effect on the value of inventions, implying that good science also leads to good technology~\citep{Poege2019science}. Moreover, in the private sector, firms perform less scientific research but instead continuously exploit discoveries from academia~\citep{Ashish2018decline, McMillan2000analysis, narin1997linkage}. Given these tight couplings between science and technology and the paucity of inquiry into the hitherto unknown process of temporal search in the scientific space, here we ask: What is the role of temporal search in the scientific space in the value of an invention? How does this role vary across technological fields? And how does it interact with temporal search in the technological space? 

Temporal search in the scientific space may play a different role in the value of inventive output. On one hand, the commercial potential of scientific knowledge tends to be uncertain~\citep{Bikard2018made}. \citet{jensen2001proofs} noted that ``when they are licensed, most university inventions are little more than a `proof of concept'". The application of newly discovered scientific knowledge in inventive process requires further trials and errors, and the reproducibility of academic studies has raised concerns regarding the usability of findings from academia \citep{Begley2012raise, Bikard2018made}. Moreover, the use of less verified knowledge not only incurs higher likelihood of inventive failures, but also affects third parties' evaluation of the invention. \citet{zhang2021patents} found that lenders are hesitant to collateralize patents associated with prior arts that are too new because of the perceived risk in determining \emph{ex ante} the patent's full potential. On the other hand, as scientific knowledge functions as public goods, old scientific knowledge may be utilized by multiple firms. Under such circumstances, the value of innovations based on past scientific knowledge is distributed across the firms developing similar technologies \citep{arora2023first}. Furthermore, scientific knowledge is updated rapidly, and failing to stay abreast of these developments may incur reinvention of the wheel. In a nutshell, the uncertainty-appropriation trade-off renders it a challenge to assert that the process of temporal search in the scientific space drives innovation outcome in the same way as in the technological space. 

To answer our research questions, we use a dataset of nearly 3.7 million patents granted at the U.S. Patent and Trademark Office (USPTO) in a 34-year period (1976-2009) to study how individual patents temporally search in the scientific space and how this type of search is related to the impact of a patent. We find that scientific search increases patent impact and doubles the odds of becoming hits. The long-established effect of technological search on patent value is dependent on scientific search: While having patent references is associated with higher citations than not having, the difference is much bigger when the patent also has scientific papers as references. Conditional on scientific search, we parameterize temporal search in the scientific space with two statistics: the mean and coefficient of variation (CV) of cited papers' ages. We find that cited science tends to be older than cited technology but has a lower variation, and there is a recent trend in referencing older science and technology. Patents that reference more recent scientific papers have more forward citations and are more likely to become hits. However, such a positive effect can be offset if patents cite papers whose ages exhibit a low dispersion. Furthermore, we find that there are positive interactions between temporal searches in the scientific and technological spaces, suggesting their mutually beneficial roles. Finally, our findings are consistent across technological fields.

\section{Literature review} \label{sec:lit}

\subsection{Typology of search}

One pivotal idea in innovation studies is that innovation results from searching and recombining prior knowledge~\citep{nelson1982role, schumpeter1939business}. The literature has proposed various types of searches and examined their performance implications. One widely studied type is local search, defined as search in the neighbor domains of an entity's current expertise. This mode of search has been shown to be a major strategy adopted by many firms \citep{helfat1994firm, stuart1996local}, and it has several advantages. \citet{kaplan2015double}, for example, found that it is more likely to be associated with patents that initiate new topics. \citet{arts2015technology} found that combining familiar knowledge in unprecedented ways is more likely to generate useful inventions, significantly increases the likelihood of breakthroughs, and reduces failure probability. \citet{jung2016quest} refined the concept of local search by emphasizing its interaction with searching original knowledge, demonstrating that originality search, when incorporated into firms' R\&D, makes local search exhibit better performances to produce high-impact breakthroughs than boundary-spanning search.

Implicit in local search is some notions of boundaries between knowledge domains, which have been delineated as technological, organizational, or geographical. \citet{rosenkopf2001beyond} considered technological and organization boundaries and found that search without spanning organizational boundaries generates lower impact patents. Focusing on the social sciences fields, \citet{schilling2011recombinant} showed that a paper's impact is associated with drawing atypical connections between different scientific fields. Similarly, \citet{kaplan2015double} found that the economic value of a patent is linked to boundary-spanning search from broader technological domains. \citet{kneeland2020exploring} identified that distant recombination contributes to the generation of outlier patents, those distant from existing inventions. \citet{castaldi2017related} found that US state-level patenting is enhanced by recombining related technologies and unrelated ones stimulate technological breakthroughs. Turning to geographical boundaries, \citet{phene2006breakthrough} found that combining domestic knowledge that is technologically distant has a curvilinear effect on breakthrough inventions, while combining international knowledge that is technologically proximate has a positive effect.

\subsection{Temporal search}

Apart from search types mentioned above, another stream of literature has focused on search along the temporal dimension, that is, recombining knowledge inputs that are produced at different points in time. This line of studies has emphasized costs and benefits of the search and the use of recent versus old knowledge. Recent knowledge opens avenues for new ideas and practices. \citet{ahuja2001entrepreneurship} found supporting evidence that experimentations with emerging technologies help firms overcome the ``maturity trap'' and create breakthrough inventions. \citet{katila2002something} revealed two aspects of search in robotics firms' creation of new products: search depth, \emph{i.e.}, the frequency of the use of existing knowledge, and search scope, the width of new knowledge. \citet{nerkar2003old} found that combining recent knowledge and knowledge with large time spans is associated with the impact of patents. In a large-scale study, \citet{mukherjee2017nearly} explored the relationship between a patent or a paper's impact and the age distribution of its references, finding that highly cited papers and patents are located in the ``hotspot'' of low mean age and high age variance. Recent studies have paid attention to moderating effects on the relationship between knowledge maturity and innovation value, pointing out the roles of technological and geographical distances \citep{capaldo2017knowledge} and the roles of firm age and size \citep{Petruzzelli2018maturity}.

Old knowledge has also been advocated as useful for knowledge creation. First, old knowledge reduces uncertainty and potential technical errors in applications \citep{turner2013strategic}, making it more reliable than newly created knowledge \citep{Lee2001recombinant}. Second, old knowledge reduces its utilization costs due to its compatibility with existing knowledge, making it easy to integrate into new applications \citep{petruzzelli2014search}. Third, a certain level of maturity is often needed to make innovation acceptable \citep{Petruzzelli2018maturity}, as new ideas without connections to current knowledge can encounter resistance to be recognized. In addition, the importance of some types of knowledge, like papers with delayed recognition \citep{ke2015defining}, can only be appreciated after sufficient amount of time when discoveries therein are brought to light due to factors like the advancement of enabling technologies \citep{cattani2006technological, nerkar2003old}.

Aged knowledge, however, may lose its advantage over new knowledge. Old knowledge that has been integrated into the space of existing knowledge quickly becomes common knowledge, and inventions that embed such knowledge become less valuable \citep{alnuaimi2016appropriability}. Moreover, old knowledge hinders individual and organizational learning \citep{ahuja2001entrepreneurship, katila2002something}, and inventors who continuously use such knowledge may fall into ``competency traps'' \citep{levitt1988organizational}, meaning that they are not able to learn superior new knowledge and practices and innovation value generated by past knowledge markedly decreases. Furthermore, the space of old knowledge lacks recombinant opportunities \citep{heeley2008recency}, leaving more room for imitation and similar ideas to competitors but less room for valuable inventions.

As mentioned before, these previous studies on temporal search of inventions are limited to searching in the technological space, leaving temporal search in the scientific space an unexplored topic, which is the focus of the present work.

\subsection{Search in the scientific space}

The role of science and technology interaction in the innovation process has long been noted by researchers \citep{sherwin1967project, price1969scientific, walsh1973technological}. Although abundant studies have investigated how technological innovations may search and rely on scientific knowledge, the literature has not yet examined temporal search in the scientific space. Just as patent references have been used to represent inventors' search activity in the technological space, scholars have used non-patent references (NPRs) that refer to scientific papers to capture the reliance of patents on science. The validity of such usage has been guaranteed by surveys. For example, \citet{roach2013lens} reported that in their survey of R\&D lab managers, NPRs better represent knowledge from public research than patent references do.

A series of empirical studies have repeatedly substantiated the reliance of technologies on science. \citet{narin1997linkage} showed that citations generated by U.S. patents to public science rapidly increased. \citet{ke2020analysis} similarly found that in biomedicine, patent-to-paper citations have been growing exponentially, doubling every 2.9 years. \citet{McMillan2000analysis} demonstrated that biotechnology firms' reliance on public research is more apparent than other industries. \citet{Ahmadpoor2017dual} devised a measure of citation distance between patents and papers to quantify the reliance of patents on science and found that fields like nanotechnology and computer science are closest to the technological space. \citet{fleming2019government} revealed that one third of U.S. patents depend on scientific research funded by the federal government.

Another line of inquiry has proved that searching in the scientific space is beneficial to inventive activities. \citet{Fleming2004science} argued that the contribution of science to technological advancement is through the alternation of the search process towards more useful combinations and empirically showed that patents referencing non-patent literature receive more and a broader scope of forward patent citations. \citet{Ahmadpoor2017dual} revealed that patents that are closer to science through citation relations are more impactful and valuable. \citet{Poege2019science} further found that the quality of scientific papers referenced by patents has a strong positive effect on the value of technological inventions, implying that good science also leads to good technology.

Taking nutrients from scientific knowledge not only promotes technological innovations but also facilitates technological breakthroughs and market values. \citet{arts2018paradise} identified that the negative effect of explorative search in the technological inventing process on breakthrough outputs can be mitigated by the reliance on science. \citet{kneeland2020exploring} found that radical patents cite more NPRs, especially more scientific papers. Finally, \citet{arora2022science} argued that inventions' reliance on science enhances markets for technology, and empirical evidence indicated that patents with citations to scientific papers are considerably more likely to be traded in intellectual property transactions.
Again, despite the heavily emphasized importance of science in technological innovations, surprisingly little theoretical or empirical work has examined how temporal search in the scientific space affects the value of innovations. Below we shall tackle this question.

\section{Data and Methods}

\subsection{Data}

We harvest patent data directly from the USPTO at its Bulk Data Storage System website (\url{https://bulkdata.uspto.gov}). We download bibliographic data files for patents granted from 1976 to 2019 and extract bibliographic information and backward citations (\emph{i.e.}, references) of these patents. The unit of our analysis is a patent, and our sample contains $3,693,101$ utility patents that are granted between 1976 and 2009, to allow each of them to have at least 10 years to accumulate forward citations.

We use a patent's backward citations to reliably capture the knowledge inputs, which is a common practice adopted in the innovation studies literature. Specifically, front-page backward citations to prior patents are used for assessing how patents rely on existing technological knowledge. Similarly, front-page NPRs that refer to scientific papers are used to assess prior scientific knowledge incorporated in the focal patent. One major difference between the two types of backward citations is that for the former, cited patents can be easily identified by patent numbers, but for the latter, only the texts of NPRs are available, without knowing if and which papers they refer to. To get the actual cited paper corresponding to a NPR, we match it with the Web of Science (WoS), a comprehensive bibliographic database for scientific papers, using the algorithm developed in \citet{ke2018comparing}, which has an accuracy of 97\%. From the WoS, we retrieve various metadata of a paper, among which the most relevant one is its publication year. Our sample of patents in total made $39,569,667$ citations to prior patents and $3,433,070$ citations to papers.

\subsection{Regression modeling}

At the patent-level, we estimate the effect of scientific search on patent's impact as
\begin{equation} \label{eq:snpr}
Y_i = \beta_0 + \beta_1 \cdot \text{has\_SNPR}_i + \gamma \cdot \text{controls}_i + \text{year}_i + \text{field}_i + \varepsilon_i \, ,
\end{equation}
where $Y_i$ is the impact of patent $i$, $\beta_0$ is the intercept, and $\beta_1$ is the coefficient of interest for whether $i$ has SNPRs. We introduce two measures to quantify impact: (1) the number of citations received by $i$ within 10 years after granting, \emph{i.e.}, forward citation count, which is a widely accepted indicator for the value of a patent~\citep{capaldo2017knowledge, kaplan2015double} and has been shown to correlate well with its technical and economic importance~\citep{Trajtenberg1990penny}; (2) a binary variable that indicates whether the focal patent is a hit, defined as its 10-year citations is among the top 5\% clustered by granted year and technological field, as operationalized as NEBR Subcategory~\citep{hall2001nber}. While 10-year forward citations provide a quantitative estimation as the extent of reuse by following-up inventions, a hit patent is extremely frequently reused compared with its peers~\citep{ahuja2001entrepreneurship, kaplan2015double, jung2016quest}. To test the robustness, we also consider forward citations until 2019 and use two other thresholds (1\% and 10\%) to identify hit patents. 

We include several control variables to account for potential confounding factors, and $\gamma$ in Eq.~\ref{eq:snpr} represents the vector of the coefficients for the controls. First, we consider whether the focal patent has patent references. Second, the broadness of its technological domains, as defined as the number of 4-digit IPC patent classes, is included to account for the tendency that technologically more broad patents may be applicable for more subsequent inventions from diverse domains. Third, we control for the number of inventors in the focal patent, as it has been shown that the size of the inventor team is positively associated with a patent's forward citations~\citep{capaldo2017knowledge, mukherjee2017nearly}. Additionally, we create a dummy variable for grant year, as there is a dependence of citations on time, and a dummy variable for technological field to account for differing rates of getting cited due to different extents of research and develop activities across fields.

In a similar fashion, the effect of temporal search in the scientific space on a patent's impact is estimated as
\begin{equation} \label{eq:temp}
Y_i = \beta_0 + \beta_1 \cdot \mu_{s,i} + \beta_2 \cdot cv_{s,i} + \gamma \cdot \text{controls}_i + \text{year}_i + \text{field}_i + \varepsilon_i \, ,
\end{equation}
where $Y_i$ is one of the two impact indicators described above. Following the characterizations of temporal search in the scientific~\citep{mukherjee2017nearly} and technological spaces~\citep{nerkar2003old}, we define the age of a cited paper as the number of years elapsed between its publication year and the granted year of the focal citing patent. We then take the mean and the coefficient of variation (CV) of the ages of all cited papers as the two measures to parametrize how the focal patent temporally searches in the scientific space, denoted as $\mu_{s,i}$ and $cv_{s,i}$. While $\mu_{s,i}$ captures the central tendency of ages, $cv_{s,i}$ quantifies the dispersion. Thus, $\beta_1$ and $\beta_2$ are the coefficients of interest. 

Control variables in Eq.~\ref{eq:temp} include mean $\mu_{t,i}$ and CV $cv_{t,i}$ of cited patents' ages, which quantify how the focal patent performs temporal search in the technological domain. The two measures control for the effects brought into play by temporal search in the technological space, as previous studies have shown that such search relates to hit inventions. Here the age of a cited patent is the number of years between its granted year and the year when the focal citing patent is granted, in line with prior works~\citep{mukherjee2017nearly, nerkar2003old}. Second, we consider the number of patent references, as the number of recombined components is positively associated with the value of an invention~\citep{kelley2013breakthroughs}. Third, we similarly include the number of cited papers~\citep{Poege2019science}. Finally, we control for academic quality of cited science, as previous studies have pointed out its role in patent value~\citep{Poege2019science, harhoff2003citations}. We measure academic quality as the average of logarithm transformed one plus number of citations in 5-years. 

Table~\ref{tab:var} reports the descriptive statistics of the constructed variables.

\section{Results} \label{sec:res}

\subsection{Search in the scientific space}

\begin{figure*}[t!]
\centering
\includegraphics[trim=0 10mm 0 0, width=\textwidth]{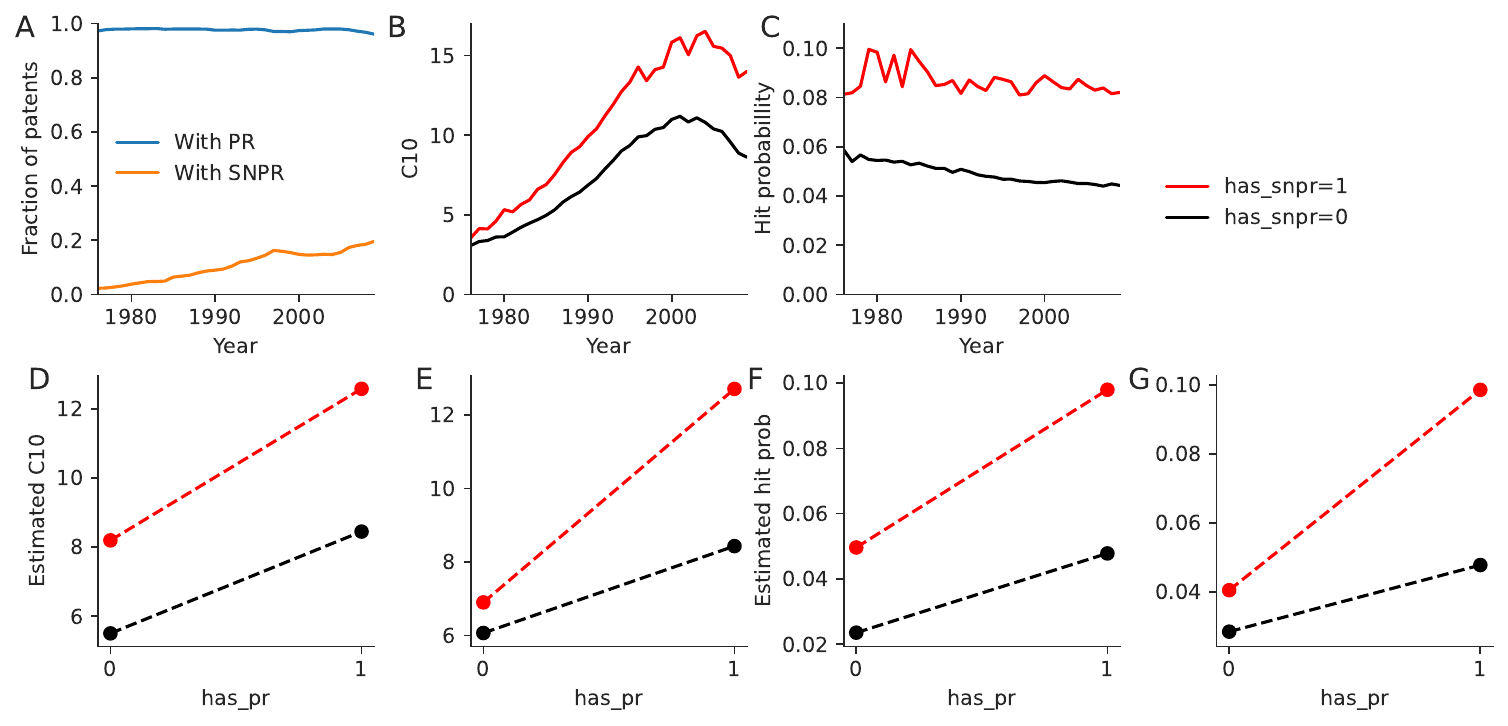}
\caption{The positive effect of search in the scientific space on patent impact. (A) Fractions of patents with patent references (PR) and SNPR over time; (B--C) Mean 10-year forward citations (B) and mean hit probabilities (C) for the groups of patents with and without SNPR; (D--E) Estimated 10-year citations based on the negative binomial regression models without (D) and with (E) considering the interaction between searches in the scientific and technological spaces; (F--G) Estimated hit probabilities based on the logistic regression models without (F) and with (G) considering the interaction between the two searches.}
\label{fig:search}
\end{figure*}

Before examining how temporal search in the scientific space affects future impact of patents, we first understand whether scientific search is associated with impact. Overall, only a minority (12.5\%) of patents have cited scientific papers, although the fraction has been increasing over time (Fig.~\ref{fig:search}A). By comparison, the vast majority (97.6\%) of patents have at least one patent reference. Relating scientific search to impact, we find that patents with SNPRs have larger future impact than those without SNPRS: They consistently collect more forward citations (Fig.~\ref{fig:search}B) and are more likely to become hit patents (Fig.~\ref{fig:search}C), as defined as their 10-year forward citations are among the top 5\% clustered by granted year and technological subfield (NEBR Subcategory) \citep{hall2001nber}. 

Further regression analyses that control for confounders confirm the advantage of scientific search in generating future impact (Table~\ref{tab:search}). After controlling for factors like search in the technological space and year and subfield fixed effects, patents with SNPRs have 49\% (${e^{0.398}-1}$) more 10-year citations (Column~1 in Table~\ref{tab:search}). Fig.~\ref{fig:search}D gives the estimated citations (predictive margins) for the four groups of patents based on whether they search in the scientific and technological spaces, indicating that patents with both patent references and SNPRs have the largest number of citations and that, conditional on citing patent reference, patents with SNPRs have mores citations than those without SNPRs. We also test whether there is a potential interaction between searches in the two spaces (Column~2 in Table~\ref{tab:search}). The significantly positive interaction term indicates that it is the case. That is, the effect of search in the technological space on forward citations is dependent on whether referencing scientific papers: While having patent references is associated with higher citations than not having patent references, the difference is much bigger when the patent also has scientific paper as reference. This dependence can be readily observed from Fig.~\ref{fig:search}E, where a steeper increase in estimated citations is apparent for patents with SNPRs. Specifically, without search in the scientific space, the effect of technological search is relatively small (39\% increase in citations); by contrast, the effect more than doubles and reaches to a 84\% increase when technological search is with the help of scientific search. 

\begingroup
\begin{table}[t!]
\centering
\caption{Regression results of patent citations and hit patents. \label{tab:search}}
\renewcommand{\arraystretch}{0.8}
\begin{tabular}{l*{8}{c}}
\hline
                    &\multicolumn{2}{c}{C10}&\multicolumn{2}{c}{Hit (5\%)}&\multicolumn{2}{c}{Hit (top 1\%)}&\multicolumn{2}{c}{Hit (top 10\%)}\\
                    & (1) & (2) & (3) & (4) & (5) & (6) & (7) & (8) \\
\hline
num\_inv            &      0.0695\sym{***}&      0.0693\sym{***}&       0.122\sym{***}&       0.122\sym{***}&       0.140\sym{***}&       0.139\sym{***}&       0.110\sym{***}&       0.110\sym{***}\\
                    &  (0.000402)         &  (0.000402)         &   (0.00124)         &   (0.00124)         &   (0.00237)         &   (0.00237)         &  (0.000960)         &  (0.000960)         \\
num\_ipc           &      0.0623\sym{***}&      0.0631\sym{***}&       0.151\sym{***}&       0.151\sym{***}&       0.212\sym{***}&       0.213\sym{***}&       0.127\sym{***}&       0.127\sym{***}\\
                    &   (0.00162)         &   (0.00162)         &   (0.00566)         &   (0.00566)         &    (0.0118)         &    (0.0118)         &   (0.00422)         &   (0.00422)         \\
has\_pr             &       0.429\sym{***}&       0.329\sym{***}&       0.736\sym{***}&       0.540\sym{***}&       0.795\sym{***}&       0.476\sym{***}&       0.685\sym{***}&       0.520\sym{***}\\
                    &   (0.00422)         &   (0.00532)         &    (0.0205)         &    (0.0268)         &    (0.0465)         &    (0.0601)         &    (0.0143)         &    (0.0185)         \\
has\_snpr           &       0.398\sym{***}&       0.128\sym{***}&       0.775\sym{***}&       0.366\sym{***}&       0.998\sym{***}&       0.355\sym{***}&       0.672\sym{***}&       0.315\sym{***}\\
                    &   (0.00203)         &   (0.00848)         &   (0.00676)         &    (0.0407)         &    (0.0139)         &    (0.0927)         &   (0.00513)         &    (0.0284)         \\
has\_pr$\times$has\_snpr&                 &       0.282\sym{***}&                     &       0.417\sym{***}&                     &       0.653\sym{***}&                     &       0.366\sym{***}\\
                    &                     &   (0.00860)         &                     &    (0.0409)         &                     &    (0.0931)         &                     &    (0.0286)         \\
Constant            &      0.0168         &       0.111\sym{***}&      -4.026\sym{***}&      -3.836\sym{***}&      -5.952\sym{***}&      -5.640\sym{***}&      -3.199\sym{***}&      -3.039\sym{***}\\
                    &    (0.0108)         &    (0.0113)         &    (0.0410)         &    (0.0443)         &    (0.0898)         &    (0.0970)         &    (0.0297)         &    (0.0318)         \\
\hline
lnalpha             &       0.211\sym{***}&       0.211\sym{***}&                     &                     &                     &                     &                     &                     \\
                    &  (0.000802)         &  (0.000802)         &                     &                     &                     &                     &                     &                     \\
\hline
Year fe             & $\checkmark$        & $\checkmark$        & $\checkmark$        & $\checkmark$        & $\checkmark$        & $\checkmark$        & $\checkmark$        & $\checkmark$        \\
Field fe            & $\checkmark$        & $\checkmark$        & $\checkmark$        & $\checkmark$        & $\checkmark$        & $\checkmark$        & $\checkmark$        & $\checkmark$        \\
\(N\)               &     3693101         &     3693101         &     3693101         &     3693101         &     3693101         &     3693101         &     3693101         &     3693101         \\
Pseudo \(R^{2}\)    &       0.033         &       0.033         &       0.016         &       0.016         &       0.020         &       0.020         &       0.014         &       0.014         \\
\textit{BIC}        &    22897190         &    22896148         &   1502420           &     1502330         &    423999           &      423964         &     2469275         &     2469125         \\
\hline
\multicolumn{5}{l}{\footnotesize Standard errors in parentheses}\\
\multicolumn{5}{l}{\footnotesize \sym{*} \(p<0.05\), \sym{**} \(p<0.01\), \sym{***} \(p<0.001\)}\\
\end{tabular}
\end{table}
\endgroup

Along a similar line, regression modeling of hit patents suggests a 117\% (${e^{0.775}-1}$) increase in the odds of becoming hits for patents referencing SNPRs, compared to comparable patents granted in the same year and in the same technological subfield but without referencing SNPRs (Column~3 in Table~\ref{tab:search}). For patents without citing prior patents, the estimated probability significantly increases from 2.3\% to 4.8\% when adding scientific references; for patents with citations to prior patents, the probability increases from 5\% to 9.8\% when citing papers (Fig.~\ref{fig:search}F). Including the interaction between searches in the two spaces reveals a boosted effect of citing patent references (Column~4 in Table~\ref{tab:search}); the increase in the estimated hit probabilities is more pronounced when the patent also cites papers (Fig.~\ref{fig:search}G). Specifically, when there is scientific search, the effect of technological search is a 144\% increase in hit probability, which is much higher than a 68\% increase when there is no scientific search. 

We perform several additional tests to ensure the robustness of our results. First, we use two other thresholds, 1\% and 10\%, to identify hit patents and find that our results are robust: Scientific search remains a significant predictor of hit patents (Columns~5--8 in Table~\ref{tab:search}). Moreover, the effect of scientific search is even more prominent for the most impactful patents (top 1\%); there is a 171\% increase in the odds of becoming top 1\% most highly-cited patents, much larger than the 117\% and 96\% increases for the top 5\% and 10\% cases, respectively. The effect of the interaction between the two searches is also stronger as we increase the threshold, suggesting that scientific search is more helpful to technological search in producing the most influential patents. Second, instead of putting a 10-year window when counting citations, we extend the window to 2019, the last year in our dataset, and find that our results still hold (Table~\ref{tab:search:ctot}). Third, to further explore whether there are cross-field heterogeneities in the effects of scientific search, we repeat our analysis separately for patents belonging to the five categories designated by NBER. These categories are Chemical, Computers \& Communications, Drugs \& Medical, Electrical \& Electronic, and Mechanical~\citep{hall2001nber}. The results indicate persistent effects of scientific search on patent impact across fields (Tables~\ref{tab:search:c10:field:1}--\ref{tab:search:c10:field:5}). 

\subsection{Temporal search and impact}

\begin{figure*}[t!]
\centering
\includegraphics[trim=0 10mm 0 0, width=\textwidth]{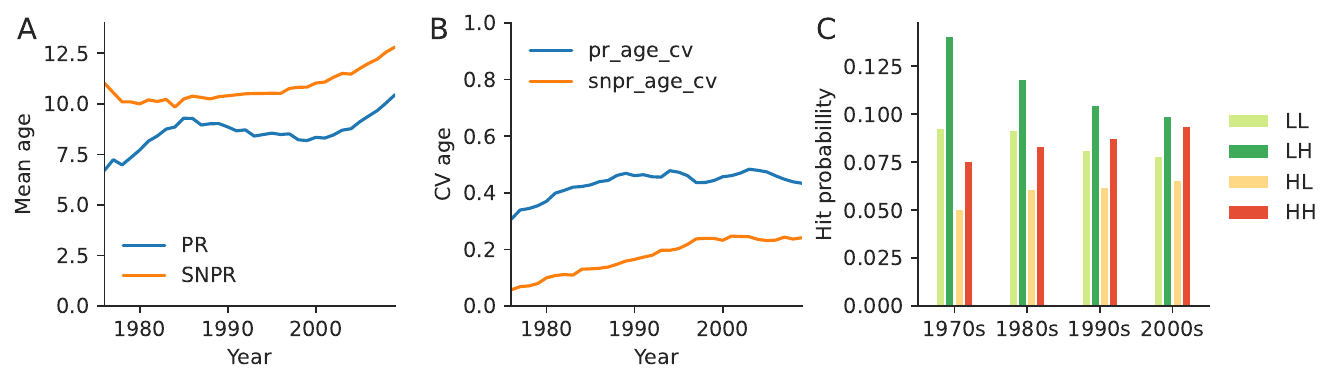}
\caption{Temporal search in the scientific space. (A) Patent-level average of $\mu_s$ and $\mu_t$, mean age of cited papers and cited patents for citing patents granted over time. (B) Patent-level average of $cv_s$ and $cv_t$, coefficient of variation of ages of cited papers and patents; (C) Hit probabilities for the four groups of patents based on whether their $\mu_s$ and $cv_s$ are below or above the respective global means over all patents in the decade.}
\label{fig:temporal}
\end{figure*}

Having established the positive linkage between patent's scientific search and future impact, we then examine how such search along the temporal dimension is associated with impact. Here we focus on the subset of patents with scientific references. Figs.~\ref{fig:temporal}A--B present patent-level statistical characterizations of temporal searches in the two spaces. We find that, for the focused subset of patents, their cited science has been constantly older than cited technology (Fig.~\ref{fig:temporal}A), which is in agreement with the view that it may take a longer time for the translation of scientific discoveries into practical applications. We also observe that, starting from around 2000, there is an upward trend of mean age for both searches, suggesting that recent patents relied on older science and technology. We postulate that this may be related to the rise of modern information technology that makes it easier for inventors to search and retrieve ``prior art'' that are more distant to the present. Turning to the dispersion of ages, cited patents exhibit a larger variation in their ages than that of cited papers (Fig.~\ref{fig:temporal}B), partly due to the fact that there are more references to patents than to papers.

Fig.~\ref{fig:temporal}C examines the relationship between temporal search in the scientific space and impact. We partition patents granted in each decade into four categories based on whether their $\mu_s$ and $cv_s$ are below (``L'') or above (``H'') the respective global means over all patents in that decade. We find that first, across all the four groups, patents have probabilities to be hits higher than the background baseline (5\%), reinforcing our previous finding that patents with SNPRs are more likely to be hits. Among the four groups, the ``LH'' one---patents located in the region of low $\mu_s$ and high $cv_s$---enjoys the largest hit probabilities (Fig.~\ref{fig:temporal}C). However, this privilege has been degrading over time. In the 1970s, the hit probability of the LH group is 2.8 times larger than that of the HL group, which decreased to 1.9, 1.7, and 1.5 times larger in the 1980s, 1990s, and 2000s, respectively. Accompanying this decrease is the increasingly higher probabilities of the HH groups, suggesting that search widely along the temporal dimension has become an important factor linked to hit patents. 

\begin{figure*}[t!]
\centering
\includegraphics[trim=0 7mm 0 0, width=\textwidth]{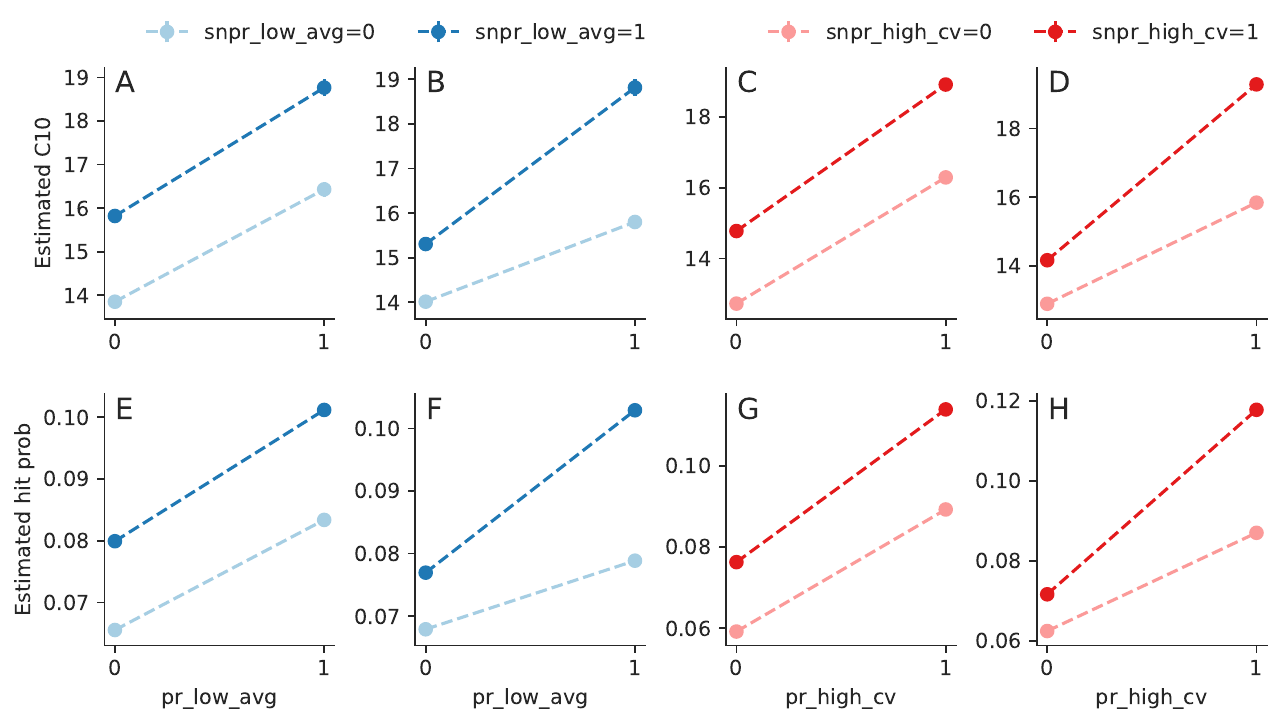}
\caption{The effects of temporal search in the scientific space on patent impact. (A--D) Estimated 10-year forward citations based on negative binomial regression models without (A, C) and with (B, D) considering the interaction between temporal searches in the scientific and technological spaces. (E--H) Estimated hit probabilities based on logistic regression models without (E, G) and with (F, H) considering the interaction.}
\label{fig:temporal-cat}
\end{figure*}

Regression analyses confirm these results net of controls (Table~\ref{tab:temporal-cat}). Patents with $\mu_s$ below the average have 14\% more citations (Column~1) and 25\% higher odds of becoming hit than the counterparts (Column~4); patents with $cv_s$ above the average have 16\% more citations (Column~1) and 32\% higher odds of being hit (Column~4). Figs.~\ref{fig:temporal-cat}A, E plot the estimated citations and hit probabilities for the four groups of patents based on whether $\mu_s$ and $\mu_t$ are below or above average, whereas Figs.~\ref{fig:temporal-cat}C, G focus on the estimated impact conditional on $cv_s$ and $cv_t$. In Columns~2 and 5, we test only the interaction between citing recent scientific and technological knowledge, and the results indicate a significantly positive interaction. This effect persists if we examine the roles of recency and dispersion simultaneously. Specifically, Column~3 suggests a positive interaction between the effects of search for recent knowledge in the two spaces, meaning that while relying on recent technological knowledge is associated with higher citations than relying on temporally distant one, the difference is much more prominent when the patent also relies on recent scientific knowledge, which is illustrated in Fig.~\ref{fig:temporal-cat}B. A similar conclusion can be drawn when the impact is defined as hit patents (Column~6 and Fig.~\ref{fig:temporal-cat}F). Positive interactions are also found for the effects of reliance on temporally dispersed scientific and technological knowledge, for both forward citations (Column~3 and Fig.~\ref{fig:temporal-cat}D) and hit patents (Column~6 and Fig.~\ref{fig:temporal-cat}H). Finally, changing the threshold for identifying hit patents has no influence in our results (Table~\ref{tab:temporal-hit}).

\begingroup
\begin{table}[!t]
\centering
\caption{Regression modeling of temporal search and impact.}
\label{tab:temporal-cat}
\renewcommand{\arraystretch}{0.7}
\begin{tabular}{l*{6}{c}}
\hline
                    & \multicolumn{3}{c}{C10} & \multicolumn{3}{c}{Hit (5\%)} \\
                    & (1) & (2) & (3) & (4) & (5) & (6) \\
\hline
num\_inv     &      0.0372\sym{***}&      0.0373\sym{***}&      0.0371\sym{***}&      0.0737\sym{***}&      0.0738\sym{***}&      0.0735\sym{***}\\
            &  (0.000964)         &  (0.000964)         &  (0.000964)         &   (0.00251)         &   (0.00251)         &   (0.00251)         \\
num\_ipc    &     0.00851\sym{*}  &     0.00822\sym{*}  &     0.00909\sym{*}  &      0.0321\sym{**} &      0.0320\sym{**} &      0.0340\sym{**} \\
            &   (0.00385)         &   (0.00385)         &   (0.00385)         &    (0.0113)         &    (0.0113)         &    (0.0113)         \\
num\_pr      &     0.00675\sym{***}&     0.00671\sym{***}&     0.00666\sym{***}&     0.00652\sym{***}&     0.00647\sym{***}&     0.00641\sym{***}\\
            & (0.0000688)         & (0.0000688)         & (0.0000688)         &  (0.000108)         &  (0.000108)         &  (0.000108)         \\
num\_snpr    &     0.00596\sym{***}&     0.00589\sym{***}&     0.00594\sym{***}&      0.0112\sym{***}&      0.0111\sym{***}&      0.0112\sym{***}\\
            &  (0.000162)         &  (0.000162)         &  (0.000162)         &  (0.000400)         &  (0.000400)         &  (0.000401)         \\
snpr\_avg\_c5 &       0.110\sym{***}&       0.111\sym{***}&       0.111\sym{***}&       0.189\sym{***}&       0.190\sym{***}&       0.191\sym{***}\\
            &   (0.00402)         &   (0.00402)         &   (0.00402)         &    (0.0120)         &    (0.0120)         &    (0.0120)         \\
pr\_low\_avg&       0.171\sym{***}&       0.118\sym{***}&       0.120\sym{***}&       0.272\sym{***}&       0.165\sym{***}&       0.169\sym{***}\\
            &   (0.00420)         &   (0.00628)         &   (0.00628)         &    (0.0124)         &    (0.0196)         &    (0.0196)         \\
pr\_high\_cv&       0.247\sym{***}&       0.249\sym{***}&       0.206\sym{***}&       0.459\sym{***}&       0.462\sym{***}&       0.370\sym{***}\\
            &   (0.00397)         &   (0.00397)         &   (0.00505)         &    (0.0121)         &    (0.0121)         &    (0.0160)         \\
snpr\_low\_avg&       0.133\sym{***}&      0.0862\sym{***}&      0.0882\sym{***}&       0.224\sym{***}&       0.136\sym{***}&       0.141\sym{***}\\
            &   (0.00406)         &   (0.00582)         &   (0.00582)         &    (0.0121)         &    (0.0173)         &    (0.0173)         \\
snpr\_high\_cv&       0.149\sym{***}&       0.149\sym{***}&      0.0938\sym{***}&       0.282\sym{***}&       0.280\sym{***}&       0.152\sym{***}\\
            &   (0.00427)         &   (0.00427)         &   (0.00588)         &    (0.0123)         &    (0.0123)         &    (0.0193)         \\
pr\_low\_avg $\times$ snpr\_low\_avg&                     &      0.0886\sym{***}&      0.0864\sym{***}&                     &       0.171\sym{***}&       0.166\sym{***}\\
            &                     &   (0.00793)         &   (0.00793)         &                     &    (0.0240)         &    (0.0240)         \\
pr\_high\_cv $\times$ snpr\_high\_cv&                     &                     &       0.103\sym{***}&                     &                     &       0.197\sym{***}\\
            &                     &                     &   (0.00752)         &                     &                     &    (0.0228)         \\
Constant    &       0.259\sym{***}&       0.279\sym{***}&       0.300\sym{***}&      -3.462\sym{***}&      -3.422\sym{***}&      -3.374\sym{***}\\
            &    (0.0323)         &    (0.0323)         &    (0.0323)         &    (0.0934)         &    (0.0935)         &    (0.0937)         \\
\hline
lnalpha     &       0.376\sym{***}&       0.376\sym{***}&       0.376\sym{***}&                     &                     &                     \\
            &   (0.00218)         &   (0.00218)         &   (0.00218)         &                     &                     &                     \\
\hline
Decade fe           & $\checkmark$        & $\checkmark$        & $\checkmark$        & $\checkmark$        & $\checkmark$        & $\checkmark$        \\
Field fe            & $\checkmark$        & $\checkmark$        & $\checkmark$        & $\checkmark$        & $\checkmark$        & $\checkmark$        \\
\(N\)       &      461377         &      461377         &      461377         &      461377         &      461377         &      461377         \\
pseudo \(R^{2}\)&       0.037         &       0.037         &       0.037         &       0.055         &       0.055         &       0.056         \\
\textit{BIC}&   3141879           &   3141768           &   3141594           &    254000           &    253962           &    253900           \\
\hline
\multicolumn{5}{l}{\footnotesize Standard errors in parentheses}\\
\multicolumn{5}{l}{\footnotesize \sym{*} \(p<0.05\), \sym{**} \(p<0.01\), \sym{***} \(p<0.001\)}\\
\end{tabular}
\end{table}
\endgroup

The above analyses use dichotomized versions for the variables of temporal search in the two spaces, we also use the original continuous variables (Table~\ref{tab:temporal}). The results show that a lower $\mu_s$ and a higher $cv_s$ are associated with a larger number of citations (Column~1) and a higher probability to be hits (Column~4), suggesting that high impact patents tend to recombine recent science but at the same time recombine temporally dispersed scientific knowledge. There are positive interactions between searching for recent scientific and technological knowledge as well as between searching for temporally dispersed knowledge sourced from the two spaces (Columns~3 and 6). These results are still robust if we change the threshold for hit patents (Table~\ref{tab:temporal-continuous-hit}). 

\begingroup
\begin{table}[t!]
\centering
\caption{Regression modeling of temporal search and impact.}
\label{tab:temporal}
\renewcommand{\arraystretch}{0.7}
\begin{tabular}{l*{6}{c}}
\hline
            & \multicolumn{3}{c}{C10} & \multicolumn{3}{c}{Hit (5\%)} \\
            & \multicolumn{1}{c}{(1)} & \multicolumn{1}{c}{(2)} & \multicolumn{1}{c}{(3)} & \multicolumn{1}{c}{(4)} & \multicolumn{1}{c}{(5)} & \multicolumn{1}{c}{(6)} \\
\hline
num\_inv    &      0.0385\sym{***}&      0.0385\sym{***}&      0.0383\sym{***}&      0.0754\sym{***}&      0.0754\sym{***}&      0.0752\sym{***}\\
            &  (0.000960)         &  (0.000960)         &  (0.000959)         &   (0.00251)         &   (0.00251)         &   (0.00251)         \\
num\_ipc    &      0.0350\sym{***}&      0.0348\sym{***}&      0.0357\sym{***}&      0.0454\sym{***}&      0.0448\sym{***}&      0.0462\sym{***}\\
            &   (0.00395)         &   (0.00395)         &   (0.00395)         &    (0.0116)         &    (0.0116)         &    (0.0116)         \\
num\_pr     &     0.00665\sym{***}&     0.00665\sym{***}&     0.00658\sym{***}&     0.00667\sym{***}&     0.00666\sym{***}&     0.00660\sym{***}\\
            & (0.0000670)         & (0.0000671)         & (0.0000670)         &  (0.000109)         &  (0.000109)         &  (0.000109)         \\
num\_snpr   &     0.00547\sym{***}&     0.00551\sym{***}&     0.00555\sym{***}&      0.0108\sym{***}&      0.0109\sym{***}&      0.0110\sym{***}\\
            &  (0.000159)         &  (0.000159)         &  (0.000159)         &  (0.000399)         &  (0.000399)         &  (0.000400)         \\
snpr\_avg\_c5 &       0.118\sym{***}&       0.117\sym{***}&       0.118\sym{***}&       0.184\sym{***}&       0.181\sym{***}&       0.182\sym{***}\\
            &   (0.00403)         &   (0.00403)         &   (0.00402)         &    (0.0121)         &    (0.0121)         &    (0.0121)         \\
pr\_age\_avg  &     -0.0142\sym{***}&     -0.0203\sym{***}&     -0.0204\sym{***}&     -0.0284\sym{***}&     -0.0411\sym{***}&     -0.0411\sym{***}\\
            &  (0.000341)         &  (0.000519)         &  (0.000519)         &   (0.00117)         &   (0.00178)         &   (0.00179)         \\
pr\_age\_cv   &       0.557\sym{***}&       0.564\sym{***}&       0.467\sym{***}&       0.921\sym{***}&       0.935\sym{***}&       0.788\sym{***}\\
            &   (0.00724)         &   (0.00726)         &   (0.00918)         &    (0.0200)         &    (0.0200)         &    (0.0260)         \\
snpr\_age\_avg&    -0.00851\sym{***}&     -0.0140\sym{***}&     -0.0140\sym{***}&     -0.0167\sym{***}&     -0.0285\sym{***}&     -0.0283\sym{***}\\
            &  (0.000263)         &  (0.000441)         &  (0.000441)         &  (0.000907)         &   (0.00154)         &   (0.00154)         \\
snpr\_age\_cv &       0.327\sym{***}&       0.331\sym{***}&       0.117\sym{***}&       0.621\sym{***}&       0.628\sym{***}&       0.286\sym{***}\\
            &   (0.00851)         &   (0.00851)         &    (0.0150)         &    (0.0233)         &    (0.0233)         &    (0.0455)         \\
pr\_age\_avg $\times$ snpr\_age\_avg&                     &    0.000470\sym{***}&    0.000472\sym{***}&                     &     0.00106\sym{***}&     0.00104\sym{***}\\
            &                     & (0.0000308)         & (0.0000308)         &                     &  (0.000108)         &  (0.000108)         \\
pr\_age\_cv $\times$ snpr\_age\_cv&                     &                     &       0.441\sym{***}&                     &                     &       0.596\sym{***}\\
            &                     &                     &    (0.0257)         &                     &                     &    (0.0675)         \\
Constant    &       0.399\sym{***}&       0.449\sym{***}&       0.486\sym{***}&      -2.900\sym{***}&      -2.790\sym{***}&      -2.732\sym{***}\\
            &    (0.0436)         &    (0.0437)         &    (0.0437)         &     (0.125)         &     (0.125)         &     (0.126)         \\
\hline
lnalpha     &       0.361\sym{***}&       0.360\sym{***}&       0.359\sym{***}&                     &                     &                     \\
            &   (0.00219)         &   (0.00219)         &   (0.00219)         &                     &                     &                     \\
\hline
Year fe     & $\checkmark$        & $\checkmark$        & $\checkmark$        & $\checkmark$        & $\checkmark$        & $\checkmark$        \\
Field fe    & $\checkmark$        & $\checkmark$        & $\checkmark$        & $\checkmark$        & $\checkmark$        & $\checkmark$        \\
\(N\)       &      461377         &      461377         &      461377         &      461377         &      461377         &      461377         \\
pseudo \(R^{2}\)&       0.039     &       0.039         &       0.040         &       0.060         &       0.060         &       0.061         \\
\textit{BIC}&   3135221           &     3134991         &     3134708         &      253114         &      253042         &      252977         \\
\hline
\multicolumn{5}{l}{\footnotesize Standard errors in parentheses}\\
\multicolumn{5}{l}{\footnotesize \sym{*} \(p<0.05\), \sym{**} \(p<0.01\), \sym{***} \(p<0.001\)}\\
\end{tabular}
\end{table}
\endgroup

\begin{figure*}[t!]
\centering
\includegraphics[trim=0 6mm 0 0, width=\textwidth]{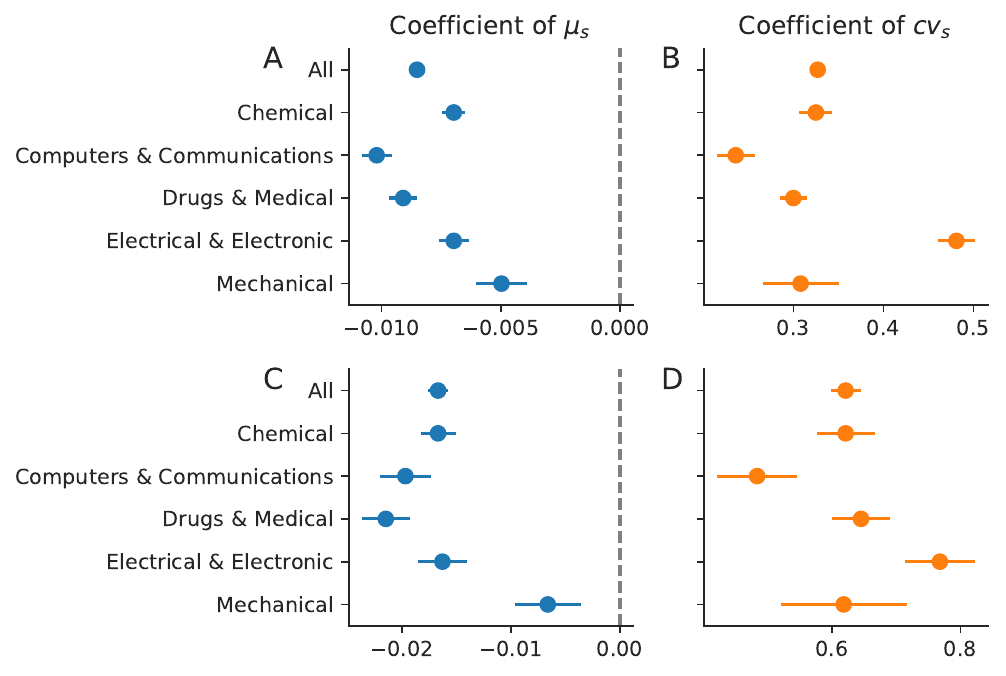}
\caption{The effects of temporal search in the scientific space on patent impact, by technological field. (A--B) Impact is based on 10-year forward citations. (C--D) Impact is based on whether patents are hits.}
\label{fig:temporal:field}
\end{figure*}

Finally, we examine whether there is field heterogeneity in the effects of temporal search in the scientific space. Subsample analyses by technological field reveal that the negative effect of mean age of cited science and the positive effect of the spread of cited science ages are persistent (Tables~\ref{tab:temporal:field:1}--\ref{tab:temporal:field:5}). However, the effect sizes vary by field (Fig.~\ref{fig:temporal:field}). The largest effect of citing recent science on patent impact is found for Computers \& Communications and Drugs \& Medical patents, reflecting the deep reliance on science of both groups of patents and rapid progresses in the fields, whereas the smallest effect is observed for mechanical patents. The effect of citing temporally diverse science is largest for Electrical \& Electronic patents and smallest for Computers \& Communications patents. 

\section{Discussion} \label{sec:dis}

The vast space of knowledge from which inventors can draw and recombine as input for new inventions has raised the questions of what their search strategies are and the performance implications. While previous inquiry has emphasized the beneficial role of search in the scientific space in technological development, it remains to be seen which and how specific types of scientific searches contribute to new inventions. Here we focus on the temporal dimension of scientific search and examine how inventions rely on prior scientific knowledge that are produced at different points in time. 

Using a large-scale sample, we have obtained three sets of results. First, patents involving scientific search have roughly 50\% more forward citations in 10 years than those without, an effect with the size comparable to the effect size of technological search. The odds of becoming hits for patents with scientific search are more than doubled than comparable patents without such search. Furthermore, there is a strong, positive interaction effect between scientific and technological searches. These results consolidate and further enrich previous studies. For example, \citet{Fleming2004science} argued science as a map in technological search, and our results provide empirical evidence for this theoretical argument. Another implication from the positive interaction between searches in the scientific and technological is from the perspective of the science side: Searching technology helps identify science with commercial potential. Second, conditional on search in the scientific space, cited science tends to be older than cited technology but has a lower variation in age, and there is a recent trend in referencing older science and technology. This represents the ``distance'' from scientific discovery to application, which needs time for commercial trials. Indeed, \citet{jensen2001proofs} reported that about 75\% of inventions were no more than a ``proof of concept'' at the time of licensing, 48\% did not have a prototype, and manufacturing feasibility was known for only 8\% of inventions. Referencing older science and technology may be attributed to the increasing complexity of technological innovations, as more knowledge elements need to be combined and some of them are from temporally ``distant'' science. Relating temporal search to patent impact, we find that relying on more recent science but at the same time referencing science with a wider spread in age are associated with more patent citations and higher likelihood of being a hit. This finding is consistent with the conclusion of temporal search in technological space, implying a general rule of knowledge search in the temporal dimension. Furthermore, we find that there are positive interactions between temporal searches in the scientific and technological spaces, suggesting their mutually beneficial roles in patent value. Third, these results are consistent across technological fields, but temporal search in the scientific space has the largest effect for the Computers \& Communications and Drugs \& Medical fields, reflecting their deep reliance on science. 

We discuss the limitations of our study for considerations of future research. First, we have looked at temporal search at the individual patent level. Future work can pay attention to firms and study how they perform temporal search in the scientific space and how it relates to the quantity and quality of their innovative output or other aspects of performances. Second, we have focused on granted patents, but inventions may take other forms like patent applications. Therefore, the generalizability of our results remains to be tested in other available data of inventive forms. Third, in light of our work and other previous large-scale studies~\citep{mukherjee2017nearly}, the roles of the age structure of prior knowledge---regardless of scientific or technological---in the impact of new knowledge seem universal, raising the questions of what possible channels are for explaining this universality and what moderators are.

\begin{acknowledgments}
Part of the work was performed when Q.K. was with Northeastern University and Syracuse University, and the data and computing resources provided there are greatly acknowledged. This work was supported in part by the National Natural Science Foundation of China (72204206 to Q.K., 72374099 to C.M.). Q.K. was additionally supported by City University of Hong Kong (Project No. 9610552, 7005968), and Hong Kong Institute for Data Science. C.M. was additionally supported by Chinese Education Department Research Foundation for Humanities and Social Sciences (19YJC870017), Jiangsu Province Social Sciences Foundation (18TQC005), and Fundamental Research Funds for the Central Universities (14380005). 
\end{acknowledgments}

\clearpage
\linespread{1.2}

\setcounter{figure}{0}
\makeatletter
\renewcommand{\thefigure}{A\@arabic\c@figure}
\makeatother

\setcounter{table}{0}
\makeatletter
\renewcommand{\thetable}{A\@arabic\c@table}
\makeatother

\setcounter{section}{0}
\makeatletter
\renewcommand{\thesection}{A\@arabic\c@section}
\makeatother

\begin{center}{\Large\textbf{Appendix}}\end{center}

\begin{table}[h!]
\centering
\caption{Summary statistics of variables.}
\label{tab:var}
\begin{tabular}{l*{8}{c}}
\toprule
Variable & Mean & Std. Dev. & Min & Max & N \\
\midrule
10-year forward citations (C10) & 8.995 & 19.02 & 0 & 3155 & 3693101\\
Total forward citations until 2019 (C2019) & 20.095 & 46.339 & 0 & 4543 & 3693101\\
\midrule
Number of inventors (num\_inv) & 2.221 & 1.611 & 1 & 76 & 3693101\\
Number of IPC patent classes (num\_ipc) & 1.132 & 0.392 & 1 & 11 & 3693101\\
Has patent reference (has\_pr) & 0.976 & 0.153 & 0 & 1 & 3693101\\
Has scientific non-patent reference (has\_snpr) & 0.125 & 0.331 & 0 & 1 & 3693101\\
Number of patent reference (num\_pr) & 10.714 & 20.135 & 0 & 1022 & 3693101\\
Number of SNPR (num\_snpr) & 0.93 & 5.474 & 0 & 170 & 3693101\\
Year & 1996.035 & 9.336 & 1976 & 2009 & 3693101\\
\midrule
Mean age of PR (pr\_age\_avg) & 8.899 & 6.287 & 0 & 134 & 461377\\
CV of PR age (pr\_age\_cv) & 0.453 & 0.299 & 0 & 2.722 & 461377\\
Mean age of SNPR (snpr\_age\_avg) & 11.28 & 7.388 & 0 & 103 & 461377\\
CV of SNPR age (snpr\_age\_cv) & 0.219 & 0.245 & 0 & 2.037 & 461377\\
Mean 5-year citations of SNPR in log (snpr\_avg\_c5) & 1.101 & 0.525 & 0 & 3.86 & 461377\\
\bottomrule
\end{tabular}
\end{table}

\clearpage

\section{Regression results of scientific search}

\begin{table}[h!]
\centering
\caption{Regression results of patent citations and hit patents. Citations are counted until 2019, and hit patents are defined accordingly. \label{tab:search:ctot}}
\begin{tabular}{l*{8}{c}}
\hline
                    &\multicolumn{2}{c}{C2019}&\multicolumn{2}{c}{Hit (5\%)}&\multicolumn{2}{c}{Hit (top 1\%)}&\multicolumn{2}{c}{Hit (top 10\%)}\\
                    &\multicolumn{1}{c}{(1)}&\multicolumn{1}{c}{(2)}&\multicolumn{1}{c}{(3)}&\multicolumn{1}{c}{(4)}&\multicolumn{1}{c}{(5)}&\multicolumn{1}{c}{(6)}&\multicolumn{1}{c}{(7)}&\multicolumn{1}{c}{(8)}\\
\hline
num\_inv            &      0.0627\sym{***}&      0.0626\sym{***}&       0.103\sym{***}&       0.102\sym{***}&       0.123\sym{***}&       0.123\sym{***}&      0.0918\sym{***}&      0.0918\sym{***}\\
                    &  (0.000407)         &  (0.000407)         &   (0.00128)         &   (0.00128)         &   (0.00248)         &   (0.00248)         &  (0.000985)         &  (0.000985)         \\
num\_ipc           &      0.0582\sym{***}&      0.0591\sym{***}&       0.132\sym{***}&       0.132\sym{***}&       0.204\sym{***}&       0.204\sym{***}&       0.113\sym{***}&       0.114\sym{***}\\
                    &   (0.00161)         &   (0.00161)         &   (0.00581)         &   (0.00581)         &    (0.0120)         &    (0.0120)         &   (0.00431)         &   (0.00431)         \\
has\_pr             &       0.440\sym{***}&       0.355\sym{***}&       0.779\sym{***}&       0.605\sym{***}&       0.867\sym{***}&       0.645\sym{***}&       0.711\sym{***}&       0.563\sym{***}\\
                    &   (0.00417)         &   (0.00526)         &    (0.0212)         &    (0.0280)         &    (0.0488)         &    (0.0655)         &    (0.0147)         &    (0.0192)         \\
\end{tabular}
\end{table}

\begin{table}[t!]
\centering
\caption{Regression results of patent citations and hit patents. Only Chemical patents are included. \label{tab:search:c10:field:1}}
\begin{tabular}{l*{8}{c}}
\hline
                    &\multicolumn{2}{c}{C10}&\multicolumn{2}{c}{Hit (5\%)}&\multicolumn{2}{c}{Hit (top 1\%)}&\multicolumn{2}{c}{Hit (top 10\%)}\\
                    & (1) & (2) & (3) & (4) & (5) & (6) & (7) & (8) \\
\hline
num\_inv             &      0.0533\sym{***}&      0.0530\sym{***}&      0.0963\sym{***}&      0.0961\sym{***}&       0.107\sym{***}&       0.106\sym{***}&      0.0840\sym{***}&      0.0838\sym{***}\\
                    &  (0.000945)         &  (0.000945)         &   (0.00294)         &   (0.00294)         &   (0.00583)         &   (0.00583)         &   (0.00226)         &   (0.00226)         \\
num\_ipc            &       0.116\sym{***}&       0.116\sym{***}&       0.235\sym{***}&       0.236\sym{***}&       0.313\sym{***}&       0.314\sym{***}&       0.215\sym{***}&       0.216\sym{***}\\
                    &   (0.00342)         &   (0.00342)         &    (0.0109)         &    (0.0109)         &    (0.0222)         &    (0.0222)         &   (0.00821)         &   (0.00821)         \\
has\_pr            &       0.395\sym{***}&       0.313\sym{***}&       0.629\sym{***}&       0.464\sym{***}&       0.680\sym{***}&       0.326\sym{**} &       0.623\sym{***}&       0.474\sym{***}\\
                    &   (0.00838)         &    (0.0107)         &    (0.0357)         &    (0.0490)         &    (0.0791)         &     (0.109)         &    (0.0256)         &    (0.0343)         \\
\end{tabular}
\end{table}

\begin{table}[t!]
\centering
\caption{Regression results of patent citations and hit patents. Only Computers \& Communications patents are included. \label{tab:search:c10:field:2}}
\begin{tabular}{l*{8}{c}}
\hline
                    &\multicolumn{2}{c}{C10}&\multicolumn{2}{c}{Hit (5\%)}&\multicolumn{2}{c}{Hit (top 1\%)}&\multicolumn{2}{c}{Hit (top 10\%)}\\
                    & (1) & (2) & (3) & (4) & (5) & (6) & (7) & (8) \\
\hline
num\_inv             &      0.0759\sym{***}&      0.0759\sym{***}&       0.121\sym{***}&       0.121\sym{***}&       0.142\sym{***}&       0.142\sym{***}&       0.116\sym{***}&       0.116\sym{***}\\
                    &  (0.000899)         &  (0.000899)         &   (0.00275)         &   (0.00275)         &   (0.00518)         &   (0.00518)         &   (0.00212)         &   (0.00212)         \\
num\_ipc            &     0.00637         &     0.00640         &    -0.00328         &    -0.00312         &      0.0614         &      0.0616         &    -0.00962         &    -0.00950         \\
                    &   (0.00430)         &   (0.00430)         &    (0.0165)         &    (0.0165)         &    (0.0344)         &    (0.0344)         &    (0.0120)         &    (0.0120)         \\
has\_pr            &       0.290\sym{***}&       0.247\sym{***}&       0.619\sym{***}&       0.496\sym{***}&       0.638\sym{***}&       0.449\sym{**} &       0.476\sym{***}&       0.395\sym{***}\\
                    &    (0.0121)         &    (0.0132)         &    (0.0601)         &    (0.0662)         &     (0.135)         &     (0.149)         &    (0.0401)         &    (0.0446)         \\
\end{tabular}
\end{table}

\begin{table}[t!]
\centering
\caption{Regression results of patent citations and hit patents. Only Drugs \& Medical patents are included. \label{tab:search:c10:field:3}}
\begin{tabular}{l*{8}{c}}
\hline
                    &\multicolumn{2}{c}{C10}&\multicolumn{2}{c}{Hit (5\%)}&\multicolumn{2}{c}{Hit (top 1\%)}&\multicolumn{2}{c}{Hit (top 10\%)}\\
                    & (1) & (2) & (3) & (4) & (5) & (6) & (7) & (8) \\
\hline
num\_inv            &      0.0400\sym{***}&      0.0396\sym{***}&      0.0695\sym{***}&      0.0693\sym{***}&      0.0794\sym{***}&      0.0792\sym{***}&      0.0625\sym{***}&      0.0623\sym{***}\\
                    &   (0.00122)         &   (0.00122)         &   (0.00357)         &   (0.00357)         &   (0.00731)         &   (0.00731)         &   (0.00271)         &   (0.00271)         \\
num\_ipc           &     -0.0348\sym{***}&     -0.0342\sym{***}&     -0.0619\sym{***}&     -0.0612\sym{***}&     0.00300         &     0.00414         &     -0.0969\sym{***}&     -0.0963\sym{***}\\
                    &   (0.00482)         &   (0.00482)         &    (0.0173)         &    (0.0173)         &    (0.0368)         &    (0.0368)         &    (0.0127)         &    (0.0127)         \\
has\_pr             &       0.524\sym{***}&       0.378\sym{***}&       0.958\sym{***}&       0.699\sym{***}&       1.139\sym{***}&       0.678\sym{***}&       0.848\sym{***}&       0.642\sym{***}\\
                    &   (0.00833)         &    (0.0143)         &    (0.0389)         &    (0.0724)         &    (0.0937)         &     (0.169)         &    (0.0265)         &    (0.0485)         \\
\end{tabular}
\end{table}

\begin{table}[t!]
\centering
\caption{Regression results of patent citations and breakthrough patents. Only Electrical \& Electronic patents are included. \label{tab:search:c10:field:4}}
\begin{tabular}{l*{8}{c}}
\hline
                    &\multicolumn{2}{c}{C10}&\multicolumn{2}{c}{Hit (5\%)}&\multicolumn{2}{c}{Hit (top 1\%)}&\multicolumn{2}{c}{Hit (top 10\%)}\\
                    & (1) & (2) & (3) & (4) & (5) & (6) & (7) & (8) \\
\hline
num\_inv            &      0.0641\sym{***}&      0.0641\sym{***}&       0.111\sym{***}&       0.111\sym{***}&       0.127\sym{***}&       0.127\sym{***}&       0.102\sym{***}&       0.102\sym{***}\\
                    &  (0.000929)         &  (0.000929)         &   (0.00300)         &   (0.00300)         &   (0.00596)         &   (0.00596)         &   (0.00229)         &   (0.00229)         \\
num\_ipc           &      0.0801\sym{***}&      0.0802\sym{***}&       0.133\sym{***}&       0.134\sym{***}&       0.153\sym{***}&       0.154\sym{***}&       0.130\sym{***}&       0.130\sym{***}\\
                    &   (0.00423)         &   (0.00423)         &    (0.0152)         &    (0.0152)         &    (0.0326)         &    (0.0326)         &    (0.0112)         &    (0.0112)         \\
has\_pr             &       0.329\sym{***}&       0.293\sym{***}&       0.617\sym{***}&       0.513\sym{***}&       0.696\sym{***}&       0.591\sym{***}&       0.609\sym{***}&       0.538\sym{***}\\
                    &    (0.0113)         &    (0.0123)         &    (0.0555)         &    (0.0627)         &     (0.128)         &     (0.147)         &    (0.0394)         &    (0.0445)         \\
\end{tabular}
\end{table}

\begin{table}[t!]
\centering
\caption{Regression results of patent citations and breakthrough patents. Only Mechanical patents are included. \label{tab:search:c10:field:5}}
\begin{tabular}{l*{8}{c}}
\hline
                    &\multicolumn{2}{c}{C10}&\multicolumn{2}{c}{Hit (5\%)}&\multicolumn{2}{c}{Hit (top 1\%)}&\multicolumn{2}{c}{Hit (top 10\%)}\\
                    & (1) & (2) & (3) & (4) & (5) & (6) & (7) & (8) \\
\hline
num\_inv             &      0.0875\sym{***}&      0.0875\sym{***}&       0.152\sym{***}&       0.152\sym{***}&       0.174\sym{***}&       0.174\sym{***}&       0.138\sym{***}&       0.138\sym{***}\\
                    &  (0.000969)         &  (0.000969)         &   (0.00294)         &   (0.00294)         &   (0.00555)         &   (0.00555)         &   (0.00230)         &   (0.00230)         \\
num\_ipc            &      0.0513\sym{***}&      0.0513\sym{***}&       0.144\sym{***}&       0.144\sym{***}&       0.185\sym{***}&       0.185\sym{***}&       0.136\sym{***}&       0.136\sym{***}\\
                    &   (0.00395)         &   (0.00395)         &    (0.0142)         &    (0.0142)         &    (0.0298)         &    (0.0298)         &    (0.0105)         &    (0.0105)         \\
has\_pr            &       0.387\sym{***}&       0.381\sym{***}&       0.733\sym{***}&       0.721\sym{***}&       0.899\sym{***}&       0.993\sym{***}&       0.617\sym{***}&       0.614\sym{***}\\
                    &    (0.0141)         &    (0.0145)         &    (0.0746)         &    (0.0786)         &     (0.181)         &     (0.205)         &    (0.0500)         &    (0.0524)         \\
\end{tabular}
\end{table}

\clearpage

\section{Regression results of temporal search}

\begingroup
\begin{table}[h!]
\centering
\caption{Regression modeling of temporal search and hit patents.}
\label{tab:temporal-hit}
\begin{tabular}{l*{6}{c}}
\hline
            & \multicolumn{3}{c}{Hit (1\%)} & \multicolumn{3}{c}{Hit (10\%)} \\
            &\multicolumn{1}{c}{(1)}&\multicolumn{1}{c}{(2)}&\multicolumn{1}{c}{(3)}&\multicolumn{1}{c}{(4)}&\multicolumn{1}{c}{(5)}&\multicolumn{1}{c}{(6)}\\
\hline
num\_inv     &      0.0832\sym{***}&      0.0832\sym{***}&      0.0828\sym{***}&      0.0647\sym{***}&      0.0647\sym{***}&      0.0646\sym{***}\\
            &   (0.00451)         &   (0.00451)         &   (0.00451)         &   (0.00202)         &   (0.00202)         &   (0.00202)         \\
num\_ipc    &      0.0589\sym{**} &      0.0588\sym{**} &      0.0614\sym{**} &      0.0185\sym{*}  &      0.0184\sym{*}  &      0.0199\sym{*}  \\
            &    (0.0217)         &    (0.0216)         &    (0.0217)         &   (0.00877)         &   (0.00877)         &   (0.00877)         \\
num\_pr      &     0.00484\sym{***}&     0.00478\sym{***}&     0.00473\sym{***}&     0.00732\sym{***}&     0.00727\sym{***}&     0.00722\sym{***}\\
            &  (0.000135)         &  (0.000135)         &  (0.000136)         &  (0.000106)         &  (0.000106)         &  (0.000106)         \\
\end{tabular}
\end{table}
\endgroup

\begingroup
\begin{table}[h!]
\centering
\caption{Regression modeling of temporal search and hit patents.}
\label{tab:temporal-continuous-hit}
\begin{tabular}{l*{6}{c}}
\hline
            & \multicolumn{3}{c}{Hit (1\%)} & \multicolumn{3}{c}{Hit (10\%)} \\
            &\multicolumn{1}{c}{(1)}&\multicolumn{1}{c}{(2)}&\multicolumn{1}{c}{(3)}&\multicolumn{1}{c}{(4)}&\multicolumn{1}{c}{(5)}&\multicolumn{1}{c}{(6)}\\
\hline
num\_inv    &      0.0857\sym{***}&      0.0856\sym{***}&      0.0854\sym{***}&      0.0664\sym{***}&      0.0664\sym{***}&      0.0663\sym{***}\\
            &   (0.00450)         &   (0.00450)         &   (0.00450)         &   (0.00203)         &   (0.00203)         &   (0.00203)         \\
num\_ipc    &      0.0773\sym{***}&      0.0762\sym{***}&      0.0775\sym{***}&      0.0302\sym{***}&      0.0296\sym{**} &      0.0306\sym{***}\\
            &    (0.0223)         &    (0.0223)         &    (0.0223)         &   (0.00906)         &   (0.00907)         &   (0.00907)         \\
num\_pr      &     0.00506\sym{***}&     0.00504\sym{***}&     0.00500\sym{***}&     0.00738\sym{***}&     0.00737\sym{***}&     0.00732\sym{***}\\
            &  (0.000137)         &  (0.000137)         &  (0.000137)         &  (0.000106)         &  (0.000106)         &  (0.000106)         \\
\end{tabular}
\end{table}
\endgroup

\renewcommand{\arraystretch}{0.7}

\begin{sidewaystable}[p]
\centering
\caption{Regression modeling of temporal search and impact. Only Chemical patents are included. \label{tab:temporal:field:1}}
\begin{tabular}{l*{8}{c}}
\hline
            &\multicolumn{2}{c}{C10}&\multicolumn{2}{c}{Hit (1\%)}&\multicolumn{2}{c}{Hit (top 5\%)}&\multicolumn{2}{c}{Hit (top 10\%)}\\
            &\multicolumn{1}{c}{(1)}&\multicolumn{1}{c}{(2)}&\multicolumn{1}{c}{(3)}&\multicolumn{1}{c}{(4)}&\multicolumn{1}{c}{(5)}&\multicolumn{1}{c}{(6)}&\multicolumn{1}{c}{(7)}&\multicolumn{1}{c}{(8)}\\
\hline
num\_inv    &      0.0400\sym{***}&      0.0400\sym{***}&      0.0700\sym{***}&      0.0701\sym{***}&      0.0718\sym{***}&      0.0718\sym{***}&      0.0590\sym{***}&      0.0590\sym{***}\\
            &   (0.00214)         &   (0.00214)         &   (0.00967)         &   (0.00967)         &   (0.00518)         &   (0.00518)         &   (0.00422)         &   (0.00423)         \\
num\_ipc    &      0.0944\sym{***}&      0.0943\sym{***}&       0.164\sym{***}&       0.162\sym{***}&       0.131\sym{***}&       0.130\sym{***}&       0.157\sym{***}&       0.156\sym{***}\\
            &   (0.00809)         &   (0.00809)         &    (0.0378)         &    (0.0378)         &    (0.0203)         &    (0.0203)         &    (0.0160)         &    (0.0160)         \\
num\_pr     &     0.00883\sym{***}&     0.00883\sym{***}&     0.00544\sym{***}&     0.00548\sym{***}&     0.00751\sym{***}&     0.00747\sym{***}&     0.00911\sym{***}&     0.00907\sym{***}\\
            &  (0.000226)         &  (0.000226)         &  (0.000358)         &  (0.000359)         &  (0.000307)         &  (0.000306)         &  (0.000309)         &  (0.000309)         \\
\end{tabular}
\end{sidewaystable}

\begin{sidewaystable}[p]
\centering
\caption{Regression modeling of temporal search and impact. Only Computers \& Communications patents are included. \label{tab:temporal:field:2}}
\begin{tabular}{l*{8}{c}}
\hline
            &\multicolumn{2}{c}{C10}&\multicolumn{2}{c}{Hit (1\%)}&\multicolumn{2}{c}{Hit (top 5\%)}&\multicolumn{2}{c}{Hit (top 10\%)}\\
            &\multicolumn{1}{c}{(1)}&\multicolumn{1}{c}{(2)}&\multicolumn{1}{c}{(3)}&\multicolumn{1}{c}{(4)}&\multicolumn{1}{c}{(5)}&\multicolumn{1}{c}{(6)}&\multicolumn{1}{c}{(7)}&\multicolumn{1}{c}{(8)}\\
\hline
num\_inv    &      0.0586\sym{***}&      0.0585\sym{***}&       0.110\sym{***}&       0.110\sym{***}&      0.0955\sym{***}&      0.0958\sym{***}&      0.0928\sym{***}&      0.0930\sym{***}\\
            &   (0.00211)         &   (0.00211)         &   (0.00909)         &   (0.00913)         &   (0.00560)         &   (0.00561)         &   (0.00462)         &   (0.00462)         \\
num\_ipc    &     -0.0333\sym{**} &     -0.0330\sym{**} &      0.0169         &      0.0231         &     -0.0572         &     -0.0552         &     -0.0585\sym{*}  &     -0.0574\sym{*}  \\
            &    (0.0110)         &    (0.0110)         &    (0.0683)         &    (0.0683)         &    (0.0369)         &    (0.0369)         &    (0.0282)         &    (0.0282)         \\
num\_pr     &     0.00703\sym{***}&     0.00697\sym{***}&     0.00661\sym{***}&     0.00650\sym{***}&     0.00858\sym{***}&     0.00848\sym{***}&     0.00922\sym{***}&     0.00912\sym{***}\\
            &  (0.000114)         &  (0.000114)         &  (0.000287)         &  (0.000288)         &  (0.000210)         &  (0.000210)         &  (0.000200)         &  (0.000200)         \\
\end{tabular}
\end{sidewaystable}

\begin{sidewaystable}[p]
\centering
\caption{Regression modeling of temporal search and impact. Only Drugs \& Medical patents are included. \label{tab:temporal:field:3}}
\begin{tabular}{l*{8}{c}}
\hline
            &\multicolumn{2}{c}{C10}&\multicolumn{2}{c}{Hit (1\%)}&\multicolumn{2}{c}{Hit (top 5\%)}&\multicolumn{2}{c}{Hit (top 10\%)}\\
            &\multicolumn{1}{c}{(1)}&\multicolumn{1}{c}{(2)}&\multicolumn{1}{c}{(3)}&\multicolumn{1}{c}{(4)}&\multicolumn{1}{c}{(5)}&\multicolumn{1}{c}{(6)}&\multicolumn{1}{c}{(7)}&\multicolumn{1}{c}{(8)}\\
\hline
num\_inv    &      0.0247\sym{***}&      0.0245\sym{***}&      0.0652\sym{***}&      0.0652\sym{***}&      0.0608\sym{***}&      0.0606\sym{***}&      0.0517\sym{***}&      0.0516\sym{***}\\
            &   (0.00162)         &   (0.00162)         &   (0.00907)         &   (0.00907)         &   (0.00454)         &   (0.00455)         &   (0.00353)         &   (0.00353)         \\
num\_ipc    &      0.0209\sym{**} &      0.0200\sym{**} &      0.0720         &      0.0713         &      0.0149         &      0.0140         &     -0.0209         &     -0.0216         \\
            &   (0.00642)         &   (0.00642)         &    (0.0419)         &    (0.0419)         &    (0.0205)         &    (0.0205)         &    (0.0154)         &    (0.0154)         \\
num\_pr      &     0.00833\sym{***}&     0.00824\sym{***}&     0.00614\sym{***}&     0.00609\sym{***}&     0.00738\sym{***}&     0.00728\sym{***}&     0.00759\sym{***}&     0.00751\sym{***}\\
            &  (0.000153)         &  (0.000154)         &  (0.000351)         &  (0.000353)         &  (0.000233)         &  (0.000234)         &  (0.000214)         &  (0.000215)         \\
\end{tabular}
\end{sidewaystable}

\begin{sidewaystable}[p]
\centering
\caption{Regression modeling of temporal search and impact. Only Electrical \& Electronic patents are included. \label{tab:temporal:field:4}}
\begin{tabular}{l*{8}{c}}
\hline
            &\multicolumn{2}{c}{C10}&\multicolumn{2}{c}{Hit (1\%)}&\multicolumn{2}{c}{Hit (top 5\%)}&\multicolumn{2}{c}{Hit (top 10\%)}\\
            &\multicolumn{1}{c}{(1)}&\multicolumn{1}{c}{(2)}&\multicolumn{1}{c}{(3)}&\multicolumn{1}{c}{(4)}&\multicolumn{1}{c}{(5)}&\multicolumn{1}{c}{(6)}&\multicolumn{1}{c}{(7)}&\multicolumn{1}{c}{(8)}\\
\hline
num\_inv    &      0.0473\sym{***}&      0.0475\sym{***}&       0.103\sym{***}&       0.103\sym{***}&      0.0867\sym{***}&      0.0868\sym{***}&      0.0756\sym{***}&      0.0756\sym{***}\\
            &   (0.00249)         &   (0.00249)         &    (0.0130)         &    (0.0130)         &   (0.00684)         &   (0.00684)         &   (0.00552)         &   (0.00552)         \\
num\_ipc    &      0.0292\sym{*}  &      0.0291\sym{*}  &      0.0597         &      0.0603         &      0.0104         &      0.0107         &     -0.0216         &     -0.0219         \\
            &    (0.0130)         &    (0.0130)         &    (0.0773)         &    (0.0774)         &    (0.0393)         &    (0.0393)         &    (0.0312)         &    (0.0312)         \\
num\_pr      &     0.00390\sym{***}&     0.00386\sym{***}&     0.00227\sym{***}&     0.00222\sym{***}&     0.00346\sym{***}&     0.00342\sym{***}&     0.00480\sym{***}&     0.00480\sym{***}\\
            &  (0.000157)         &  (0.000157)         &  (0.000354)         &  (0.000355)         &  (0.000259)         &  (0.000260)         &  (0.000275)         &  (0.000275)         \\
\end{tabular}
\end{sidewaystable}

\begin{sidewaystable}[p]
\centering
\caption{Regression modeling of temporal search and impact. Only Mechanical patents are included. \label{tab:temporal:field:5}}
\begin{tabular}{l*{8}{c}}
\hline
            &\multicolumn{2}{c}{C10}&\multicolumn{2}{c}{Hit (1\%)}&\multicolumn{2}{c}{Hit (top 5\%)}&\multicolumn{2}{c}{Hit (top 10\%)}\\
            &\multicolumn{1}{c}{(1)}&\multicolumn{1}{c}{(2)}&\multicolumn{1}{c}{(3)}&\multicolumn{1}{c}{(4)}&\multicolumn{1}{c}{(5)}&\multicolumn{1}{c}{(6)}&\multicolumn{1}{c}{(7)}&\multicolumn{1}{c}{(8)}\\
\hline
num\_inv    &      0.0300\sym{***}&      0.0301\sym{***}&      0.0700\sym{***}&      0.0702\sym{***}&      0.0534\sym{***}&      0.0536\sym{***}&      0.0683\sym{***}&      0.0686\sym{***}\\
            &   (0.00487)         &   (0.00487)         &    (0.0184)         &    (0.0184)         &    (0.0115)         &    (0.0115)         &   (0.00976)         &   (0.00977)         \\
num\_ipc    &      0.0290         &      0.0292         &      -0.102         &      -0.103         &      0.0162         &      0.0169         &      0.0263         &      0.0269         \\
            &    (0.0185)         &    (0.0185)         &    (0.0834)         &    (0.0834)         &    (0.0463)         &    (0.0463)         &    (0.0385)         &    (0.0385)         \\
num\_pr     &     0.00696\sym{***}&     0.00698\sym{***}&     0.00495\sym{***}&     0.00494\sym{***}&     0.00595\sym{***}&     0.00599\sym{***}&     0.00588\sym{***}&     0.00595\sym{***}\\
            &  (0.000209)         &  (0.000209)         &  (0.000409)         &  (0.000409)         &  (0.000374)         &  (0.000376)         &  (0.000375)         &  (0.000376)         \\
\end{tabular}
\end{sidewaystable}

\end{document}